\documentclass[letter]{aa} 

\usepackage{graphicx}
\usepackage{natbib}
\usepackage{color}
\usepackage{upgreek}
\usepackage{soul}

\newcommand{\teff}{T_{\rm eff}}
\newcommand{\kms}{km\,s$^{-1}$}

\newcommand{\vsini}{v \sin i}

\usepackage{txfonts}
\begin{document} 

\titlerunning{Magnetic field of HD\,169142}
\authorrunning{Hubrig et al.}

\title{The enigmatic magnetic field of the planet hosting\\Herbig Ae/Be star HD\,169142}
   \author{
	  S.~Hubrig\inst{1}
          \and
          S.~P.~J\"arvinen\inst{1}
          \and
          I.~Ilyin\inst{1}
          \and
          M.~Sch\"oller\inst{2}
                     }

   \institute{Leibniz-Institut f\"ur Astrophysik Potsdam (AIP),
              An der Sternwarte~16, 14482~Potsdam, Germany\\
              \email{shubrig@aip.de}
           \and European Southern Observatory, Karl-Schwarzschild-Str.~2, 85748~Garching, Germany
}

   \date{Received MM DD, 2024; accepted MM DD, 2024}

 
  \abstract
  {
    Recent observations of the accretion disk around the Herbig Ae/Be star HD\,169142 revealed its complex and asymmetric
    morphology indicating the presence of planets.
        The knowledge of the  magnetic field structure in host stars is indispensable for our understanding
    of the magnetospheric interaction between the central stars, the circumstellar (CS) environment, and planetary companions.
    }
   {
     We intend to study the geometry of the magnetic field of HD\,169142.
}
{
 We measured the mean longitudinal magnetic field from high resolution ESPaDOnS and HARPS\-pol spectra 
of HD\,169142 using the Least Square Deconvolution technique.
Additionally, the spectral variability of hydrogen lines is studied using dynamical spectra. 
   }
     {Our analysis of the Stokes~$V$ spectra reveals the presence of definitely detected narrow Zeeman features observed using
line masks with neutral iron lines. On two observing epochs, we also obtain marginally detected broad Zeeman features. 
To explain the simultaneous appearance of narrow and broad Zeeman features,
we discuss different scenarios, including one scenario related to a non-photospheric origin
of the narrow Zeeman features due to magnetospheric interaction with warm CS matter.
In an environment such as a wind or an accretion disk, spectral lines may
form over a relatively large volume, and the field topology may therefore be complex not only in latitude and azimuth,
but in radius as well.
Dynamical plots of the H$\beta$ line show an intriguing very complex structure with 
appearing and disappearing absorption features, which can be related to the complex morphology of the CS
matter with asymmetric dust clump structures. The profiles of spectral lines
belonging to different elements are variable, indicating the presence of chemical spots.
}
   {}

 \maketitle
%

\section{Introduction}
\label{sect:intro}

The motivation to search for the presence of a magnetic field and its structure in the rather bright ($m_{\rm V}=7.7$)
A9-F0 Herbig Ae star 
HD\,169142 at an estimated age of 6$^{+6}_{\rm -3}$\,Myr \citep{Grady2007}
arises from the recent detections of several signatures typical for giant planet formation in its face-on protoplanetary disk
(e.g., \citealt{Pohl2017,Bertrang2018,Toci2020}).
HD\,169142 is located at a distance of 114\,pc \citep{GuzmanDiaz2021} and is surrounded by several ring-like structures
(e.g.\ \citealt{Law2023}),  spirals
(e.g.\ \citealt{Gratton2019}), and asymmetric dust clump structures at the innermost ring
(e.g.\ \citealt{Per2019}).
Numerous studies of the complex and asymmetric morphology of the accretion disk around HD\,169142
considered that many of the accretion disk structures originate from planet-disk interactions.
The presence of giant planets having a few Jupiter masses, one located inside the inner dust
cavity at a distance of less than 20\,au and another giant planet in the gap between the two dust rings (between 35 and 55\,au),
has been suggested to explain the appearance of two prominent dust rings (e.g., \citealt{Fedele2017,Mac2019,Per2019}). 
Using a re-reduction of previous SPHERE YJH band observations, \citet{Hammond2023} recently confirmed 
the detection of a compact source in the face-on protoplanetary disk surrounding HD\,169142.
To match the position of the observed compact source, the authors included in their model
two giant planets with $M\sim2.5\,M_ J$, one located in the inner cavity, at 17\,au,
and one located between the two rings, at a position angle of 44$^{\circ}$. 

While the kinematic pattern and line profiles studied using observations with the
Atacama Large Submillimeter Array are consistent with an outer disk seen at 13$^{\circ}$ inclination from face-on orientation
\citep{Raman2006,Panic2008}, the inner circumstellar (CS) disk with a radius
of $2.2\pm0.6$\,au has been found to be misaligned between $10-23^{\circ}$ with respect to the outer disk
(e.g.\ \citealt{Francis2020}).
According to \citet{Poblete2022}, the morphology
of the disk structure can be represented by a central misaligned CS disk and four more distant rings 
at 26, 57, 64 and 77\,au, respectively.
To explain the misalignment of the inner disc and the azimuthal concentration
of spiral arms at the innermost ring at a distance of 26\,au observed by \citet{Poblete2019},
the authors suggested the presence of an inner stellar binary and a circumbinary planet.
Their 3D hydrodynamical simulations of circumbinary disks using the PHANTOM smoothed particle hydrodynamics
code \citep{Price2018a} indicate that a companion with a mass ratio of 0.1,
with a semimajor axis of 9.9\,au, an eccentricity of 0.2, and an inclination of 90$^{\circ}$,
together with a $2\,M_{J}$ coplanar planet on a circular orbit at 45\,au reproduce well both,
the structures at the innermost ring observed at 1.3\,mm and the shape of the spiral features in scattered light observations.
On the other hand, the suggested binarity scenario for HD\,169142 is not supported by available
spectroscopic \citep{Corporon1999} and interferometric observations \citep{Kobus2020}, nor by Gaia data \citep{Vioque2018}.
 
In view of the reported presence of planets in the system HD\,169142,
a study of the stellar magnetic properties of HD\,169142 is particularly valuable for our understanding of the
possible magnetic star-planet interaction. Through the stellar wind and the magnetic field,
stars usually create a magnetized plasma environment extending outwards into the surrounding space (e.g.\ \citealt{Strugarek2015}).
Since optical ultranarrowband imagery of HD\,169142 revealed  the presence  of H$\alpha$ emission \citep{Grady2007},
this star is expected to possess a magnetosphere \citep{Sundqvist2012}.
One of the most common approaches to investigate the interaction between a magnetic star and a planet is to search for non-thermal
chromospheric \ion{Ca}{ii}\,H and K excess emissions.
These lines are not in emission in the spectra of HD\,169142, but, as we show in Fig.~\ref{afig:ca} in the Appendix:
the line cores are strongly variable
and are possibly filled-in by emission.
\citet{Grady2007} identified HD\,169142 as a X-ray and UV line emission source,
but no radio monitoring campaign to search for flares has been carried out so far.

As of today, about two dozen Herbig Ae/Be stars have been reported to possess large-scale organized magnetic fields
with a mean longitudinal magnetic field strength between 0.1 and 1.2\,kG
(e.g., \citealt{Hubrig2009,Hubrig2015}). Among the magnetic Herbig Ae/Be stars,
HD\,101412 possesses the strongest surface magnetic field ever measured in any Herbig Ae/Be star,
of up to 3.5\,kG \citep{Hubrig2010}.
The few Herbig Ae/Be stars with intriguing hints of planets have not yet been studied
in detail using spectropolarimetry.
\citet{Alecian2013}
used four spectropolarimetric observations of HD\,169142
recorded in 2005 between February 20 and August 24 using the Echelle SpectroPolarimetric Device
for the Observation of Stars \citep[ESPaDOnS;][]{Donati2006} installed at the Canada-France-Hawaii Telescope (CFHT) and
reported a non-detection with an upper limit of about $50-100$\,G.

To characterise the magnetic field of HD\,169142, we recently acquired additional high-resolution
polarimetric spectra using the High Accuracy Radial
velocity Planet Searcher polarimeter \citep[HARPS\-pol;][]{Harps}
attached to the European Southern Observatory (ESO) 3.6\,m telescope.
This paper is laid out as follows: Section~\ref{sect:obs} describes the obtained observations,their reduction,
and the measurement procedure. 
In Section~\ref{sect:disc}, we consider the distinct variability of
line profiles belonging to different elements and discuss possible scenarios to explain the observed magnetic field structure.

\section{Observations and magnetic field measurements}
\label{sect:obs}

The reduced ESPaDOnS spectra, which are now publically available in the CFHT
archive, have a wavelength coverage from 3700 to 10\,480\,\AA{}
and a spectral resolution of $R\approx65\,000$.
We complemented the ESPaDOnS observations with four high resolution ($R\approx110\,000$) HARPS\-pol observations obtained
during our visitor run in 2022 April--May (Prg.~ID~0109.C-0265(A)).
HARPS\-pol has a wavelength coverage from 3780 to 6910\,\AA{},
with a small gap between 5259 and 5337\,\AA{}.
The data were reduced on La Silla using the HARPS\-pol data reduction pipeline.
The normalization of the spectra to the continuum level is described in detail by \citet{Hubrig2013}.
The assessment of the longitudinal magnetic field measurements 
is presented in our previous papers (e.g.\ \citealt{Hubrig2018}, \citealt{Jarvinen2020}). 
Similar to the work of \citet{Alecian2013} and our previous studies, to increase the 
signal-to-noise ratio ($S/N$) by a multiline approach, we employed the least-squares deconvolution (LSD) technique. 
The details of this technique, 
also the description how the LSD Stokes~$I$, Stokes~$V$, and diagnostic null spectra are calculated, were 
presented by \citet{Donati1997}.
The fundamental parameters of HD\,169142 have recently been
reported in several studies: $\log\,g=4.20\pm0.25$ and
$\vsini=55\pm0.8$\,\kms{} have been reported by \citet{Saffe2021} and the effective temperature $\teff=7250\pm125$\,K, the mass 
$M=1.55^{+0.03}_{\rm -0.00}$\,$M_{\sun}$, and the radius $R=1.51\pm0.05\,R_{\sun}$  are listed in the study by \citet{GuzmanDiaz2021}.

\begin{table*}
\caption{
Logbook of observations and results of magnetic field measurements for the Herbig Ae star HD\,169142.
\label{tab:obsall}
}
\centering
\begin{tabular}{cccrccc r@{$\pm$}ll}
\hline
\multicolumn{1}{c}{Instr.} &
\multicolumn{1}{c}{JD} &
\multicolumn{1}{c}{Exposure} &
\multicolumn{1}{c}{$S/N$} &
\multicolumn{1}{c}{Line} &
\multicolumn{1}{c}{FAP}&
\multicolumn{1}{c}{Det.} &
\multicolumn{2}{c}{$\left< B \right>_{\rm z}$} &
\multicolumn{1}{c}{Remarks} \\
\multicolumn{1}{c}{} &
\multicolumn{1}{c}{(245\,0000$+$)} &
\multicolumn{1}{c}{time (s)} &
\multicolumn{1}{c}{} &
\multicolumn{1}{c}{mask} &
\multicolumn{1}{c}{} &
\multicolumn{1}{c}{flag} &
\multicolumn{2}{c}{(G)} \\
\hline
E  & 3422.1630 & 2400 &121 & \ion{Fe}{i}, \ion{Si}{ii} & $1.6\times10^{-4}$ & MD & $-$1310 & 118& broad feature \\
E  & 3424.1403 & 2400 & 97 & \ion{Fe}{i}, \ion{Si}{ii} & \multicolumn{1}{c}{} & ND & \multicolumn{2}{c}{}& too noisy\\
E  & 3512.9597 & 2400 &144 &  \ion{Fe}{i}              & $3.5\times10^{-6}$ & DD & 123 & 17& crossover-like\\
E  & 3606.8105 & 2000 &239 & \ion{Fe}{i}               & $>10^{-3}$ & ND & \multicolumn{2}{c}{}& broad feature? \\
H  & 9694.8707 & 4800 &145 & \ion{Fe}{i}               &  $<10^{-10}$    & DD & 19   & 8 \\
H  & 9697.7998 & 6400 &140 & \ion{Fe}{i}               &  $8.5\times10^{-7}$ & DD & $-$42  & 15 & \\
H  & 9697.7998 & 6400 &145 & \ion{Cr}{ii}, \ion{Fe}{ii} &  $6.3\times10^{-4}$ & MD & $-$58  & 41 & broad feature\\
H  & 9698.9001 & 4000 &150 & \ion{Fe}{i}               & $2.8\times10^{-7}$ & DD & $-$92 & 17& \\
H  & 9699.8618 & 4000 &210 & \ion{Fe}{i}               & $3.8\times10^{-7}$  & DD & $-$46 & 17& possible blend? \\
\hline
\end{tabular}
\tablefoot{
The first column gives the name of the spectropolarimeter used -- with E for ESPaDOnS and H for HARPS\-pol.
The second column presents the Julian Date at the middle of the exposure, while in the third and fourth columns 
we show the corresponding exposure times and the signal-to-noise ratio measured in the Stokes~$I$ spectra
in the spectral region around 4510\,\AA{}.
The line mask used, the false alarm probability (FAP) values, the detection flag --
where DD means definite detection, MD marginal detection, and ND no detection --,
the measured LSD mean longitudinal magnetic field strength,
and a remark on the Stokes~$V$ feature(s), are presented in Columns~5--9.
}
\end{table*}

\begin{figure*}
\centering 
\includegraphics[width=0.640\textwidth]{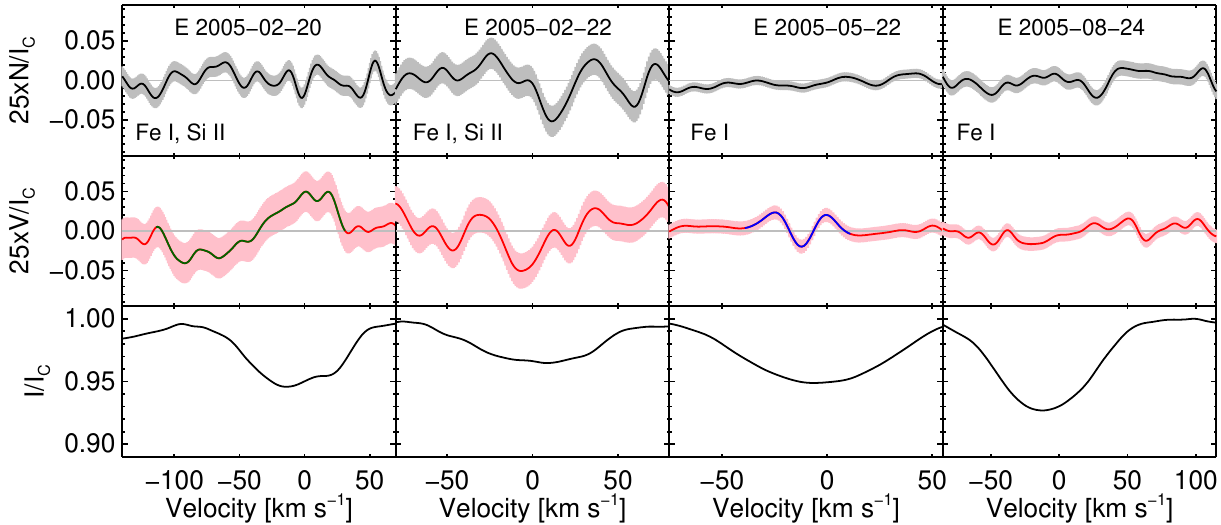}
\includegraphics[width=0.640\textwidth]{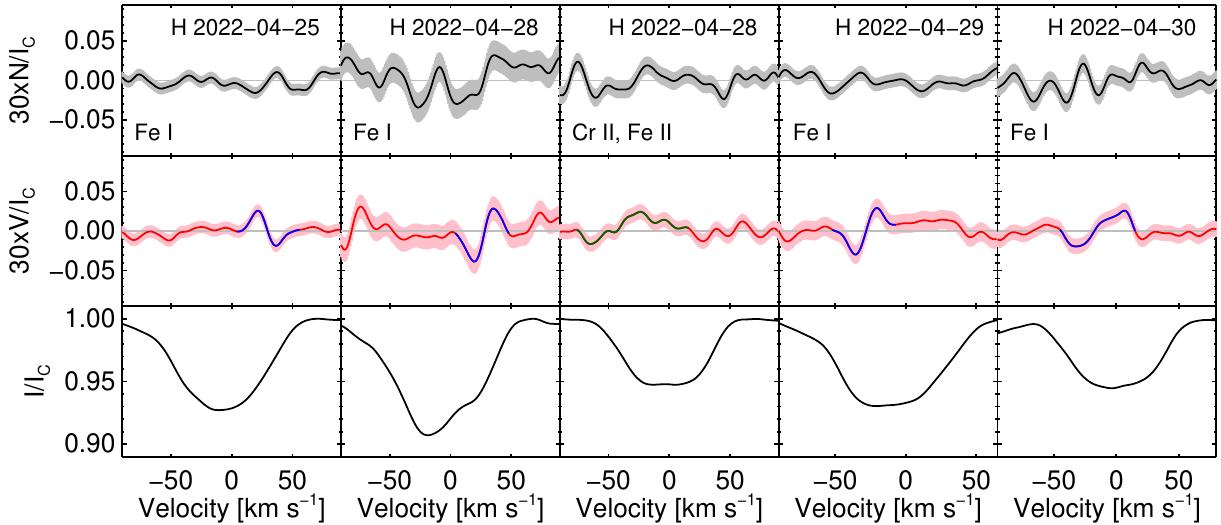} 
\caption{
LSD Stokes~$I$, $V$, and diagnostic null $N$ spectra (from bottom to top) of HD\,169142.
These are calculated using ESPaDOnS (bottom) and HARPS\-pol (top) observations. The employed line masks and
the dates of observations are indicated.
The  Stokes~$V$ and $N$ spectra are magnified by a factor of 30 for better visibility.
The rather narrow Zeeman feature identified in the Stokes~$V$ spectra as definite detections are highlighted in blue,
whereas the broad Zeeman features identified as marginal detections are highlighted in green.
The grey bands in the $N$ spectra and the red bands in the Stokes~$V$
spectra correspond to a 1$\sigma$ uncertainty.
}
\label{fig:meas}
\end{figure*}

The spectrum of HD\,169142 is very line-rich, exhibiting numerous \ion{Fe}{i} lines. Only lines that appear to be unblended or minimally
blended in the Stokes~$I$ spectra were included in the line mask.
The resulting LSD profiles are scaled according to the
line strength and sensitivity to the magnetic field. 
To evaluate, whether the detected Zeeman features are spurious or definite detections, 
we followed the generally adopted procedure to use the false alarm probability (${\rm FAP}$) based on 
reduced $\chi^{2}$ test statistics \citep{Donati1992}:
the presence of a Zeeman feature
is considered as a definite detection (DD)
if ${\rm FAP} \leq 10^{-5}$,
as a marginal detection (MD) if  $10^{-5}<{\rm FAP}\leq 10^{-3}$,
and as a non-detection (ND) if ${\rm FAP}>10^{-3}$.

Previous studies of magnetic Herbig Ae/Be stars revealed
that different elements typically show different abundance distributions across the stellar surface, and therefore
sampling the magnetic field in different manners (e.g., \citealt{Hubrig2010,Jarvinen2016}).
Thus, the geometry of the magnetic field
can potentially be studied by measurements using spectral lines of each element separately.
As an example, we show in Fig.~\ref{afig:hd98} the difference in the strength
of the longitudinal magnetic fields obtained 
for the magnetic Herbig Ae/Be star HD\,98922
(e.g., \citealt{Hubrig2013,Jarvinen2019a}) using exclusively the \ion{Ti}{ii} line mask and the line mask
containing in addition \ion{Cr}{ii} and
\ion{Fe}{ii} lines. Therefore, different line masks have been used in the LSD analysis of HD\,169142.

In Table~\ref{tab:obsall} we
summarise the results of the magnetic field measurements of HD\,169142, providing additional information on the Julian date at mid-exposure,
the $S/N$ measured at 4510\,\AA{}, and the instrument used. In the last column of this table we list our
remarks related to the appearance of the Zeeman features detected in the Stokes~$V$ spectra.
The LSD Stokes~$I$, $V$, and diagnostic null $N$ spectra for each observation are presented in Fig.~\ref{fig:meas}.
Using the line mask with \ion{Fe}{i} lines, our measurements show definite detections for all four HARPS\-pol observations and
one ESPaDOnS observation, with longitudinal magnetic field strengths up to 120\,G.
The definite detection in the ESPaDOnS observation from 2005 May 22 shows a Zeeman crossover-like feature.
Such features are usually explained by the presence of magnetic fields of mixed polarities
over the visible hemisphere and the rotational Doppler effect in fast rotating stars.
No detection using this line mask was obtained for the  ESPaDOnS observation in 2005 August 24. 
The distinct rather narrow Zeeman features with FAP-values
corresponding to definite detections are in most cases shifted relative to the centre
of the underlying Stokes~$I$ profiles, suggesting that the structure of the magnetic field
can be complex with magnetic/chemical spots distributed over the stellar surface.
For the remaining two ESPaDOnS observations, we employed a line mask containing \ion{Fe}{i} and \ion{Si}{ii} lines. 

Surprisingly, we detect broad Zeeman
features on two different observing epochs: one in the ESPaDOnS observation acquired in 2005 February 20
with $\left< B \right>_{\rm z}=-1310\pm118$\,G obtained using the line mask containing \ion{Fe}{i} and \ion{Si}{ii} lines,
and another one in the HARPS\-pol observations in 2022 April 28 with
$\left< B \right>_{\rm z}=-58\pm41$\,G obtained using a line mask containing \ion{Cr}{ii} and \ion{Fe}{ii} lines.
Both broad Zeeman features highlighted in Fig.~\ref{fig:meas} in green colour correspond
to marginal detections with  ${\rm FAP}=1.6\times10^{-4}$ and  ${\rm FAP}=6.3\times10^{-4}$, respectively.
For the same  HARPS\-pol observations from 2022 April 28, using a line mask containing \ion{Fe}{i} lines, the observed Zeeman feature
is much more narrow, but shows the same (negative) polarity.
It is possible that a weak broad Zeeman feature is also present in the Stokes~$V$ spectrum in the observations from 2005 August 24 and
as a blend with the narrow Zeeman feature in the observation acquired in 2022 April 29.
However, higher quality spectra with much higher $S/N$ are necessary to confirm their appearance.

Inspecting the LSD plots in Fig.~\ref{fig:meas}, we also realize that the diagnostic $N$ spectra obtained for the HARPS\-pol observation in 
2022 April 28 and the ESPaDOnS observation in 2005 February 22
display large irregular structures. These observations have been obtained with the
lowest $S/N$. On the other hand, we suspect that the reason for such an appearance, also for other distinct smaller structures
observed in the $N$ spectra calculated for few other observations, is related  to the extremely complex CS environment,
which is variable on very short time scales.
Since spectropolarimetric exposure times are extremely long,
the observations could have been significantly affected by short-time CS variability.
  
\section{Discussion}
\label{sect:disc}

\begin{figure*}
\centering 
\includegraphics[width=0.200\textwidth]{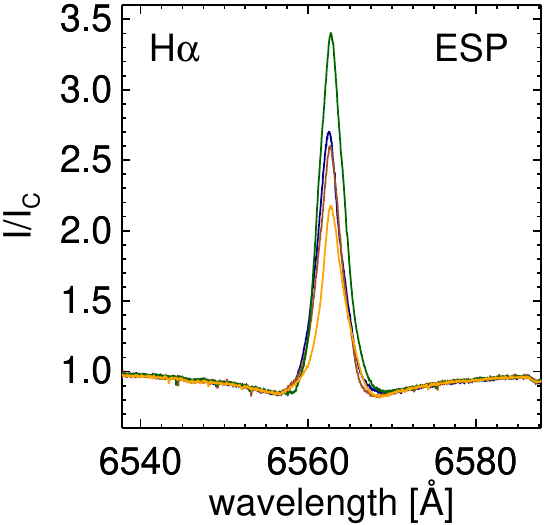}
\includegraphics[width=0.200\textwidth]{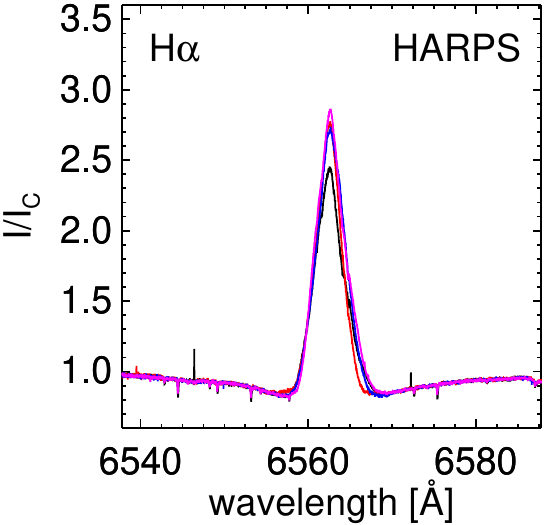} 
\includegraphics[width=0.200\textwidth]{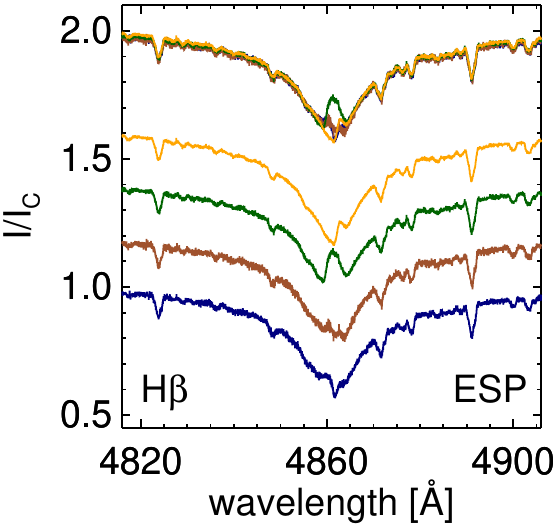}
\includegraphics[width=0.200\textwidth]{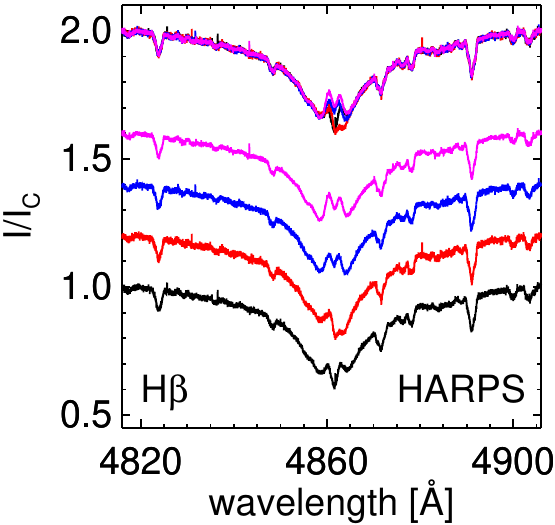} 
\caption{
  ESPaDOnS and HARPS\-pol overplotted spectral profiles.
H$\alpha$ profiles are presented in the first and second panels on the left.
  The other two panels display individual H$\beta$ line profiles in chronological order
(oldest at bottom) and the overplotted profiles on the top.
 }
\label{fig:alpha}
\end{figure*}

\begin{figure*}
\centering 
\includegraphics[width=0.378\textwidth]{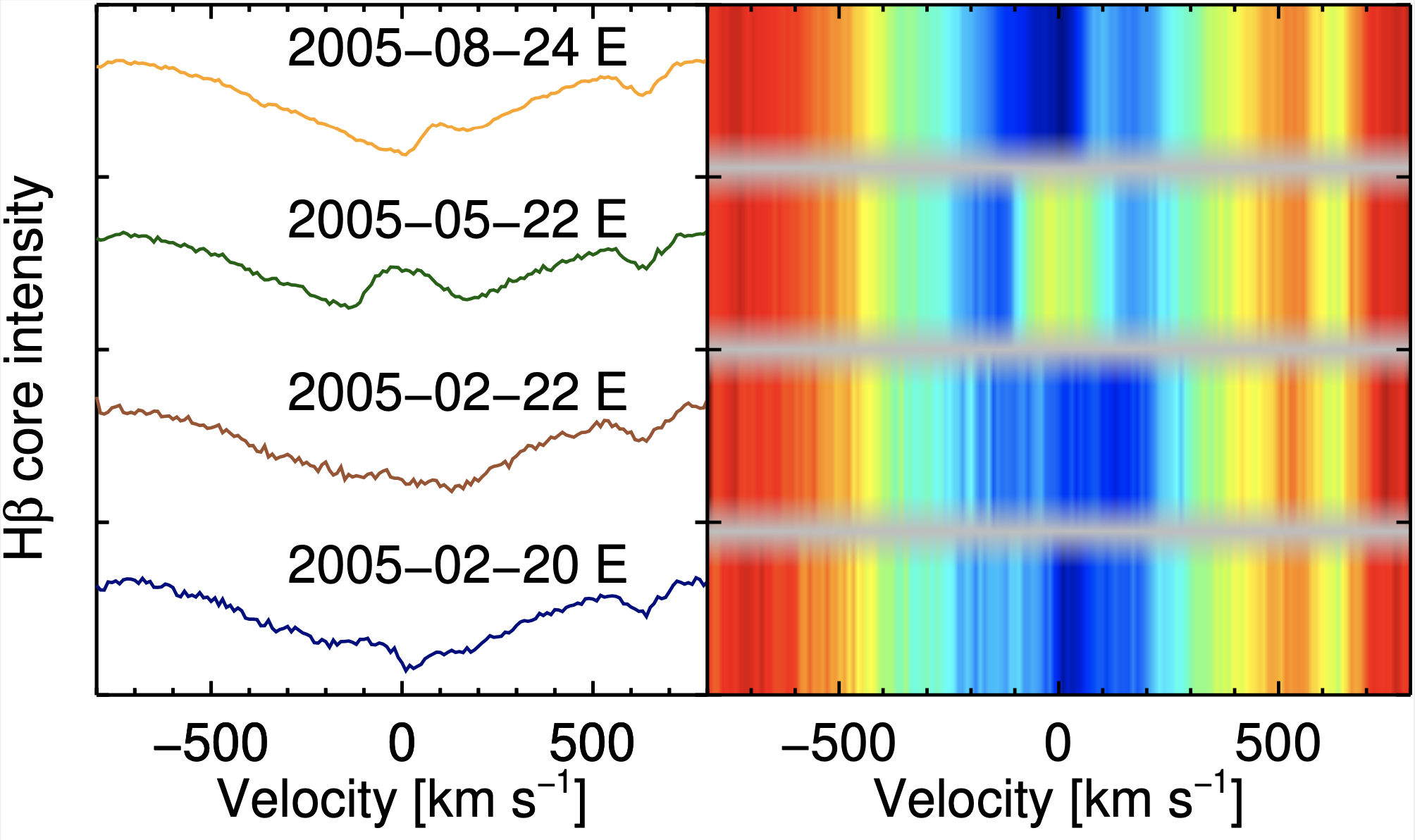}
\includegraphics[width=0.378\textwidth]{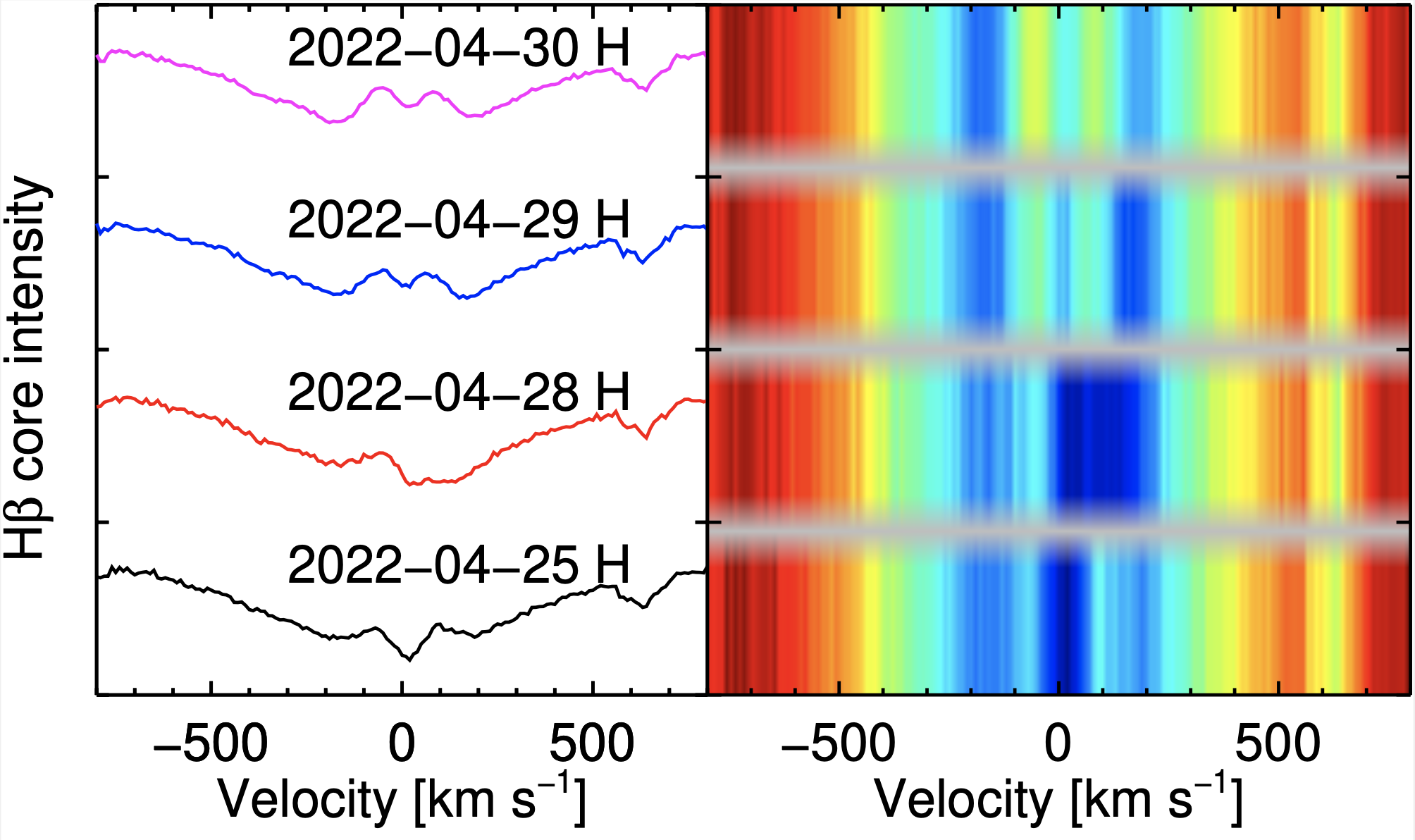}
\includegraphics[width=0.230\textwidth]{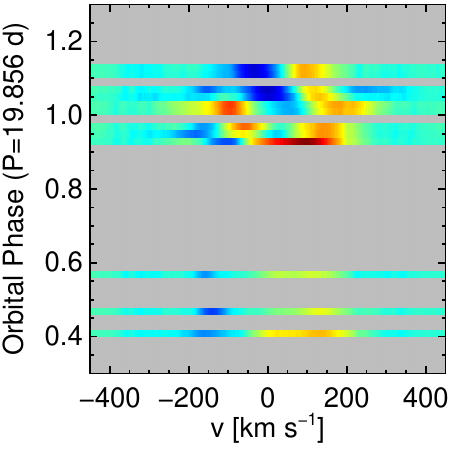}
\caption{
Dynamical spectra.
{\it Left and Middle:} Strongly variable cores of H$\beta$ line profiles in HD\,169142 and corresponding dynamical spectra
constructed using ESPaDOnS and HARPS\-pol spectra.
{\it Right:} Dynamical spectrum calculated for the variable emission
H$\beta$ line observed in the SB2 system HD\,104237 consisting of a weakly
magnetic Herbig Ae/Be star and a strongly magnetic T\,Tauri component. The
ring-like features are centred at the phases where the maximum longitudinal
magnetic field is observed.}
\label{fig:dyn}
\end{figure*}

It is striking that the narrow definitely detected Zeeman features are observed exclusively for line masks containing
\ion{Fe}{i} lines, whereas the inclusion of lines belonging to ions
(\ion{Si}{ii}, \ion{Cr}{ii}, or \ion{Fe}{ii}) produces
broad Zeeman features on two epochs, albeit identified as marginal detections.
As mentioned in several studies in the past, a strong contamination by CS matter is observed in the spectra of cool Herbig Ae/Be stars
with numerous lines of neutral iron (e.g., \citealt{Hubrig2009,Jarvinen2015}). The study of \citet{Jarvinen2015} of the strongly
magnetic Herbig Ae/Be star HD\,101412 showed that the shape
and the position of the LSD Stokes~$V$ profiles obtained using a line mask with neutral Fe lines can differ from the
position of the Stokes~$V$ profiles obtained with masks for iron ions. This has been explained by the contamination caused
by surrounding warm CS matter in the form of wind or accretion.
While \ion{Fe}{i} Stokes~$V$ profiles
calculated for HD\,101412 with an accretion disk inclined by $80\pm7^{\circ}$ \citep{Fedele2008},
i.e.\ viewed close to edge-on, appear  shifted to the blue with respect to the Stokes~$I$ profiles, the \ion{Fe}{i} Stokes~$V$
profiles observed for HD\,169142 show both shifts to the blue and to the red, and one Stokes~$V$ profile with a crossover shape. Different
radial velocity shifts of \ion{Fe}{i} of these profiles are probably caused by the different inclination angle of the
stellar rotation axis of HD\,169142, which is viewed not so far from pole-on.

A heavy obscuration by the CS matter can be studied using the line shapes of the hydrogen lines in the HARPS\-pol and ESPaDOnS
spectra. Plots with H$\alpha$ and H$\beta$ profiles for all observing epochs are presented
in Fig.~\ref{fig:alpha}. The variability of the core of the H$\beta$ profiles appears especially intriguing as the central part of the core
displays two emission peaks in the HARPS\-pol observations from 2022 April 29 and 30. The central emission is also visible in the
ESPaDOnS observations from 2005 May 22, but appears unresolved probably due to the lower spectral resolution. The dynamical plots
of the cores of the H$\beta$ lines presented in the left and middle panels of Fig.~\ref{fig:dyn} show a very complex structure with
a number of appearing and disappearing absorption features that can be related to the complex morphology of the CS
matter with asymmetric dust clump structures. The distance between the emission peaks in 
velocity space is larger for the observation on April 30, accounting for 135\,km\,s$^{-1}$, with the red-shifted position of the dip
between the emission peaks accounting for 20\,km\,s$^{-1}$. 
The strongest emission in the H$\alpha$ profiles corresponds to the observing epochs
where the emission peaks in the core of the H$\beta$ lines are observed.

It is possible that both emission peaks coming into sight in 2022 April 29 and 30 and
highlighted as yellow central strips in Fig.~\ref{fig:dyn}
can be related to stellar magnetospheres typically observed in dynamical spectra constructed for hydrogen lines in
magnetic upper-main sequence stars (e.g.\ \citealt{Kueker2024}). Notably, a first spectroscopic imprint of a magnetosphere
displaying a ring-like feature centred at the phase where the maximum magnetic field is observed has 
recently been presented for the the nearly face-on Herbig Ae/Be star HD\,190073 \citep{Jarvinen2025}.
Also the dynamical spectrum constructed for the observed H$\beta$ lines in the nearly face-on SB2 system HD\,104237,
consisting of a Herbig Ae/Be component and a T\,Tauri component,
displays a ring-like feature centred at the phase corresponding
to the maximum magnetic field in the strongly magnetic T\,Tauri component 
(see the right side of Fig.~\ref{fig:dyn}).
Because of the reported nearly face-on disk orientation in HD\,169142, we actually expected to discover a similar ring-like feature for
this star too.

To characterize the variability of the H$\beta$ line cores on short-time scales, we also downloaded from the ESO archive
two HARPS high cadence time series
acquired in 2008 March 17 and March 21 in the framework of the ESO Prg.~ID~080.C-0712(A). The dynamical spectra for these series, each
over about two hours, are presented in Fig.~\ref{afig:dyn2008}. The observed changes in the line cores, the change in the widths and separation
of the emission peaks over 131\,min on March~21 and small changes in the absorption features over 129.6\,min on March 17 seem to confirm
that HD\,169142 is a fast rotator with a CS matter morphology appearing variable already on short time scales.

Taking into account that the measured longitudinal
magnetic field is aspect dependent and changes with rotation phase, to ascertain the structure of the magnetic field, it is necessary
to sample different rotation phases. The rotation period of HD\,169142 is
however unknown. Using $\vsini=55\pm0.8$\,\kms{} \citep{Saffe2021},
the radius $R=1.51\,R_{\sun}$ \citep{GuzmanDiaz2021}, and assuming that the inclination of the stellar rotation axis is the same as
the inclination of the outer disks seen at 13$^{\circ}$,
we obtain a high equatorial velocity of 244\,km\,s$^{-1}$ and a rotation period of about 0.313\,d.
For the misalignment of $10-23^{\circ}$ suggested by \citet{Francis2020}, the rotation period can be as long as 0.816\,d.

Considering the strong aspect dependence of the measured longitudinal magnetic field
and the heavy obscuration by the CS matter clearly detected in the dynamical spectra,
we can only speculate that 
the observed broad Zeeman features might be due to the presence of a globally organized  kG-order photospheric magnetic field,
whereas narrow Zeeman features observed for the \ion{Fe}{i} line mask are of non-photospheric origin related to the 
magnetospheric interaction with warm CS matter.
On the other hand, in the case of the fast rotating cool F0-type
star HD\,169142, it is possible that it still retains a shallow convective zone and that the observed enhanced X-ray and UV line
emission may reflect rotational enhancement of chromospheric and transition region activity \citep{Grady2007}.
As a result, small-scale magnetic structures
would not have Ohmically decayed yet due to the young age. As the star evolves,
it could acquire the more typical dipole-dominated magnetic field like that of Ap/Bp stars.
Admittedly, the issue whether cool Herbig Ae/Be stars exhibit small-scale magnetic fields associated with temperature spots
typical to those observed in late-type stars is currently unexplored.

In Appendix~\ref{sect:profiles} in Figs.~\ref{afig:h_elem}
and \ref{afig:h_elemdyn} we show that the line profiles
belonging to different elements are variable and appear slightly split at some epochs,
indicating a possible presence of surface
chemical spots. The presence of chemical spots has previously been reported also for some other magnetic Herbig Ae/Be
stars (e.g.\ \citealt{Jarvinen2019b,Hubrig2010}).

The motion of the CS gas in pre-main sequence stars is expected to be governed by the magnetic field. In contrast to T\,Tauri stars,
both the field origin and the way it couples to the CS matter are not clear for Herbig Ae/Be stars. 
While magnetic surveys of Herbig Ae/Be stars so far mainly targeted the detection of magnetic fields, it is important to understand
their three-dimensional structure with respect to the stellar axis and disk orientation. In a CS environment, such as a wind or
an accretion disk, spectral lines may
form over a relatively large volume, and the field topology may therefore be complex not only in latitude and azimuth, but in radius as well.
Future multi-epoch observations of Doppler-shifted spectropolarimetric contributions from photospheric and CS environmental diagnostic lines
are necessary to apply for HD\,169142 the technique of Zeeman-Doppler tomography. This will allow us to determine the
correspondence between the magnetic field structure and the radial density and temperature profiles.

\begin{acknowledgements}
We are very grateful
to the anonymous referee for their valuable comments.
Based on observations made with ESO Telescopes at the La Silla
Paranal Observatory under programme IDs~0109.C-0265(A) and 080.C-0712(A),
and on observations collected at the Canada-France-Hawaii
Telescope (CFHT), which is operated by the National Research Council of
Canada, the Institut National des Sciences de l'Univers of the Centre
National de la Recherche Scientifique of France, and the University of Hawaii.
\end{acknowledgements}

%
\bibliographystyle{aa} 
\bibliography{aa53225-24} 
%

%
%
%

\begin{appendix}

\section{Observed changes in the line profile of the \ion{Ca}{ii}\,K line}
\label{sect:ca}

\begin{figure}
\centering
\includegraphics[width=0.380\textwidth]{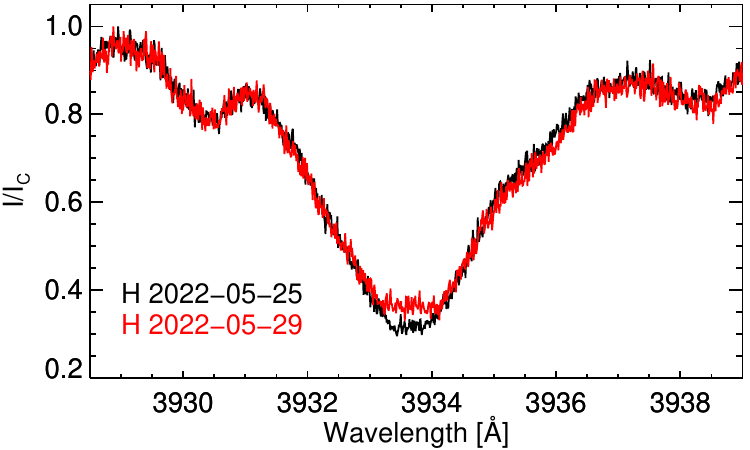}
\caption{
\ion{Ca}{ii}\,K line profiles recorded in the spectra of HD\,169142 on two different epochs.
}
\label{afig:ca}
\end{figure}

To show distinct variability of the \ion{Ca}{ii}\,K lines, we display in Fig.~\ref{afig:ca} the profile of this line
on two different epochs. The \ion{Ca}{ii}\,K lines appear in absorption, but the line cores are strongly variable
and are possibly filled-in by emission.

\section{Magnetic field strength of HD\,98022 using different line masks}
\label{sect:elem}

\begin{figure}
\centering
\includegraphics[width=0.240\textwidth]{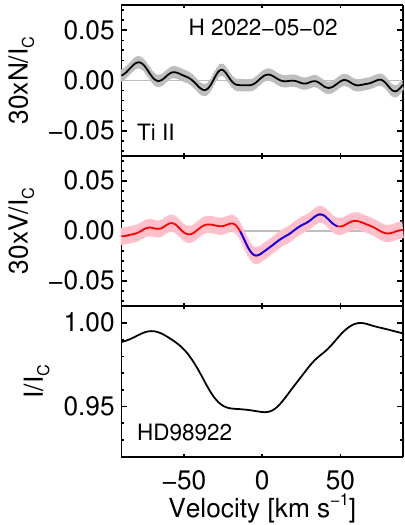}
\includegraphics[width=0.240\textwidth]{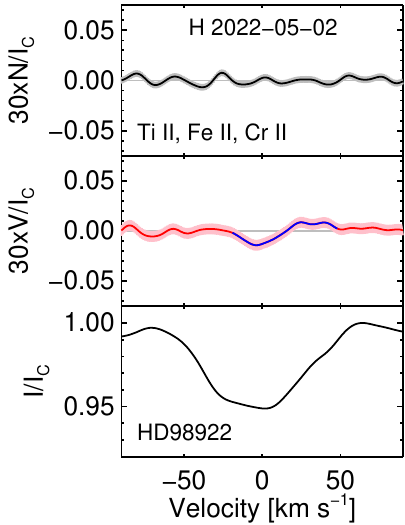}
\caption{
LSD Stokes~$I$, $V$, and diagnostic null $N$ spectra (from bottom to top)
calculated for HD\,98922 observed with HARPS\-pol on 2022 May 2.
The  Stokes~$V$ and $N$ spectra are magnified by a factor of 30 for better visibility.
Zeeman signatures identified in the Stokes~$V$ spectra are highlighted in blue.
The grey bands in the $N$ spectra and the red bands in the Stokes~$V$
spectra correspond to a 1$\sigma$ uncertainty.
}
\label{afig:hd98}
\end{figure}

In Fig.~\ref{afig:hd98} we show the difference in the appearance of 
the LSD Stokes~$V$ profiles calculated for the magnetic Herbig Ae/Be star HD\,98922
(e.g., \citealt{Hubrig2013,Jarvinen2019a}) using exclusively a \ion{Ti}{ii} line mask and a line mask
containing in addition \ion{Cr}{ii} and \ion{Fe}{ii} lines. HD\,98922  was found to be
consistent with a face-on orientation with a disk inclination of $5\pm5^{\circ}$ \citep{Ganci2024}.
For the strength of the magnetic field using exclusively \ion{Ti}{ii} lines we obtain
$\left< B \right>_{\rm z}=-211\pm25$\,G with ${\rm FAP}=8.5\times10^{-10}$, whereas the field strength is
lower,  $\left< B \right>_{\rm z}=-157\pm16$\,G,
using in addition the \ion{Cr}{ii} and \ion{Fe}{ii} lines, and has a less significant ${\rm FAP}=1.9\times10^{-8}$.
These results confirm that different elements typically show different abundance distributions across the stellar surface,
and therefore sampling the magnetic field in different manners.

\section{H$\beta$ variability in spectroscopic HARPS time series acquired in 2008}

\begin{figure}
\centering 
\includegraphics[width=0.443\textwidth]{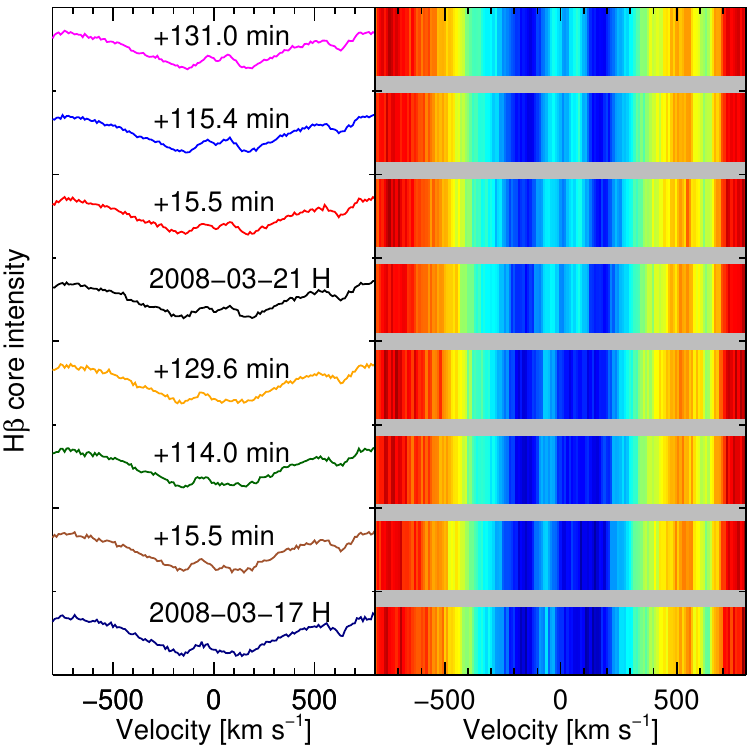} 
\caption{
Similar to Fig.~\ref{fig:dyn} but for the spectroscopic HARPS\-pol time series acquired in 2008.}
  \label{afig:dyn2008}
\end{figure}

In Fig.~\ref{afig:dyn2008} we show the H$\beta$ variability in the spectroscopic HARPS\-pol time series acquired in 2008.
The observed changes of  the intensities of the cores of the H$\beta$ lines indicates that HD\,169142 is most probably a fast rotator
with a CS matter morphology appearing variable already on short
time scales.

  \section{LSD Stokes~$I$  profile variability observed for different elements}
  \label{sect:profiles}

\begin{figure*}
\centering 
\includegraphics[width=.9\textwidth]{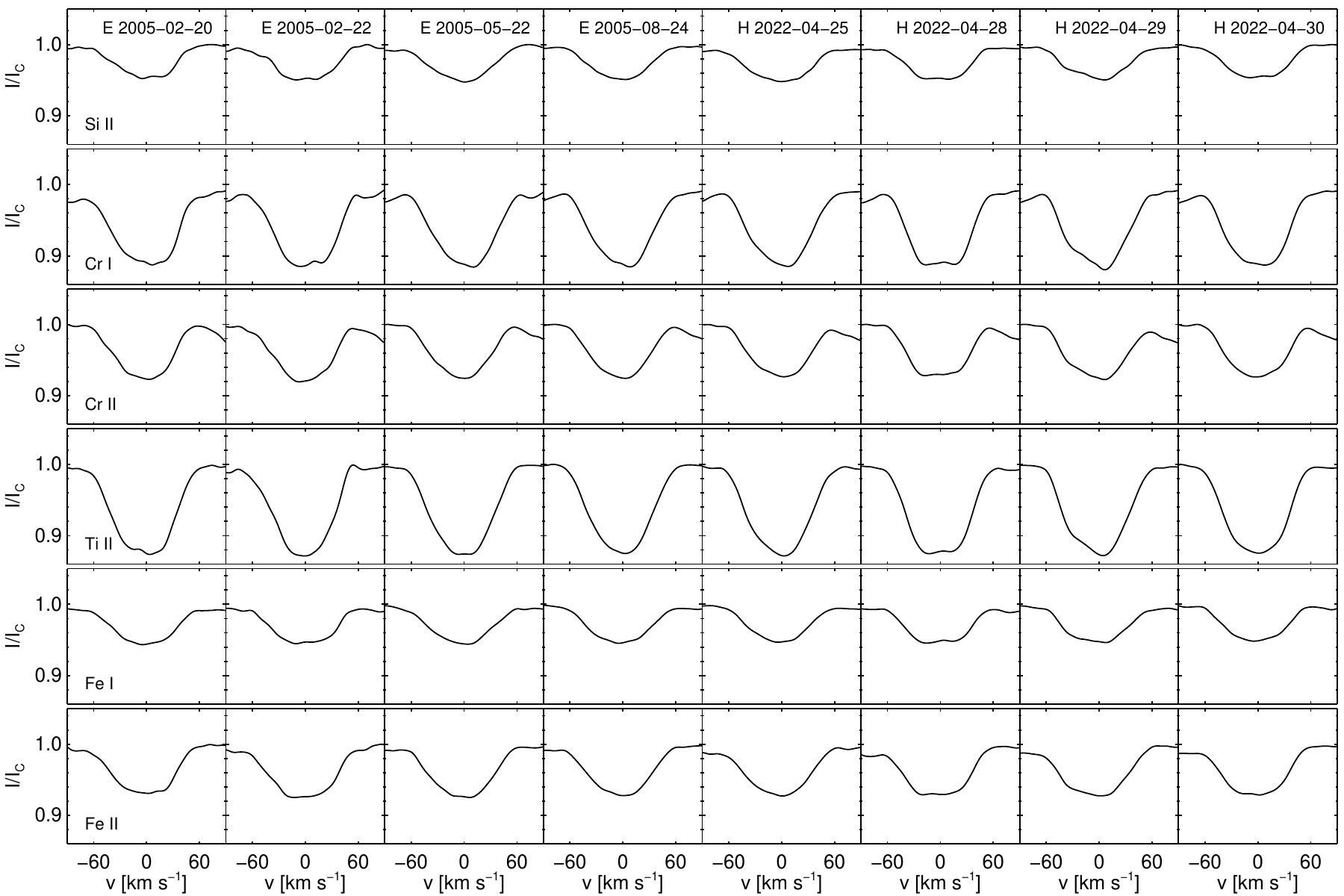}
\caption{
LSD Stokes~$I$ profiles calculated for different elements identified in HARPS\-pol and ESPaDOnS spectra.
}
\label{afig:h_elem}
\end{figure*}

\begin{figure*}
\centering
\includegraphics[width=.3\textwidth]{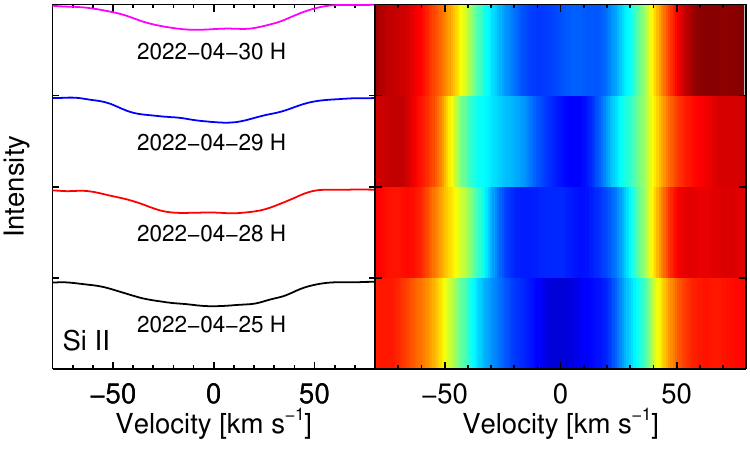}
\includegraphics[width=.3\textwidth]{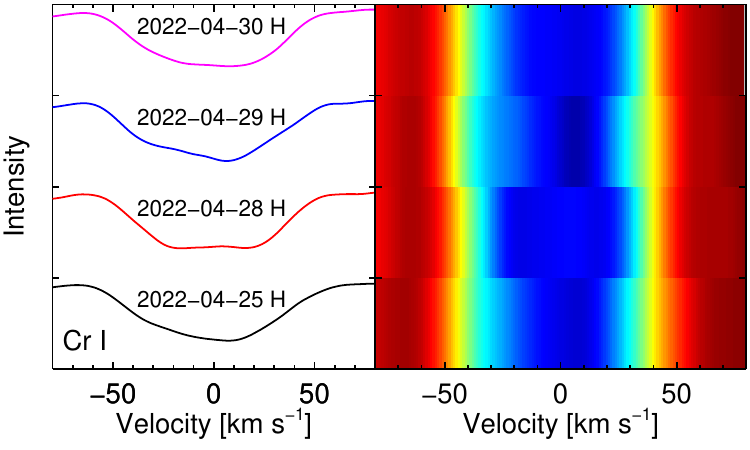}
\includegraphics[width=.3\textwidth]{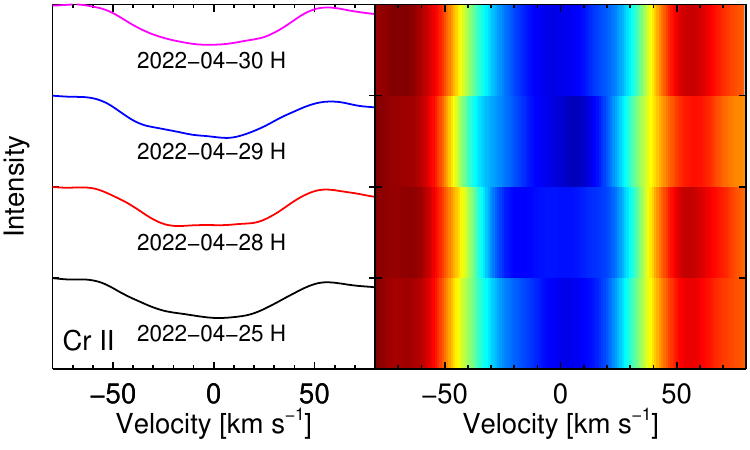}
\includegraphics[width=.3\textwidth]{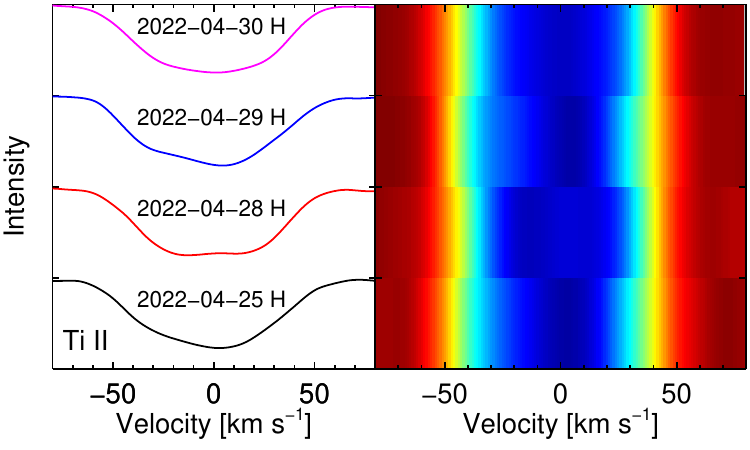}
\includegraphics[width=.3\textwidth]{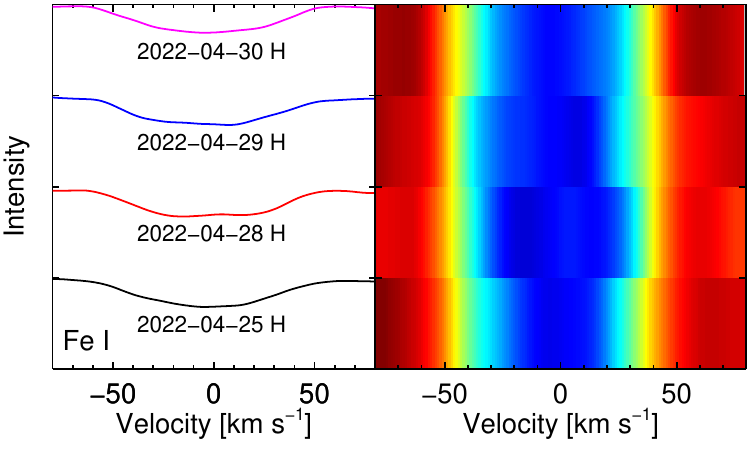}
\includegraphics[width=.3\textwidth]{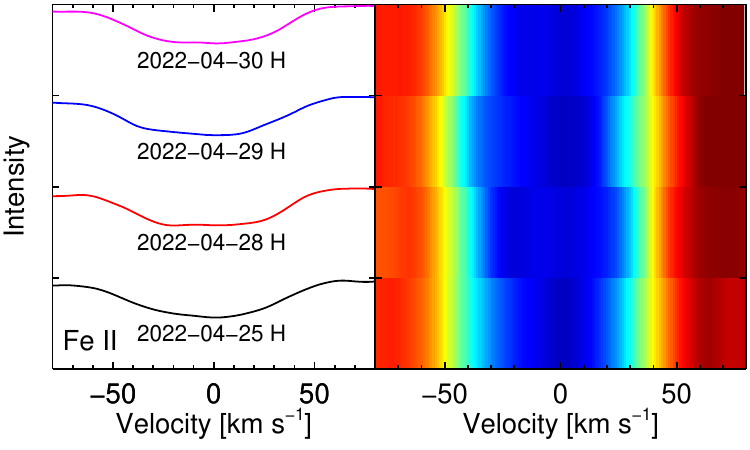}
\caption{
Dynamical spectra of the LSD Stokes~$I$ profiles calculated for different elements
identified in the HARPS\-pol observations.
The dark blue colour indicates the areas with chemical spots.
}
\label{afig:h_elemdyn}
\end{figure*}

The LSD Stokes~$I$ profiles for Si, Ti, Cr, and Fe
are calculated using moderately strong lines that are better suited to monitor the character of the variability.
As shown in Fig.~\ref{afig:h_elem}, all profiles appear variable with sometimes a slightly split structure,
most pronounced in the HARPS\-pol observations acquired on 2022 April 28. For better visibility of the line profile variability,
we also show in Fig.~\ref{afig:h_elemdyn} dynamical spectra for each element.
Due to the higher spectral resolution of the HARPS\-pol observations compared to the ESPaDOnS observations,
only the HARPS\-pol observations are presented. Variability in the line profiles belonging to different
elements is usually explained by an inhomogeneous element distribution on the surface of Herbig Ae/Be stars
(e.g.\ \citealt{Jarvinen2019b,Hubrig2010}).

 \end{appendix}

\end{document}